\documentclass[twocolumn,conference]{IEEEtran}
\usepackage[T1]{fontenc}
\usepackage[latin9]{inputenc}
\usepackage{array}
\usepackage{units}
\usepackage{textcomp}
\usepackage{multirow}
\usepackage{amsmath}
\usepackage{graphicx}
\usepackage[unicode=true,
 bookmarks=true,bookmarksnumbered=true,bookmarksopen=true,bookmarksopenlevel=1,
 breaklinks=false,pdfborder={0 0 0},pdfborderstyle={},backref=false,colorlinks=false]
 {hyperref}
\hypersetup{pdftitle={Your Title},
 pdfauthor={Your Name},
 pdfpagelayout=OneColumn, pdfnewwindow=true, pdfstartview=XYZ, plainpages=false}

\makeatletter

\providecommand{\tabularnewline}{\\}

\usepackage[caption=false,font=footnotesize]{subfig}
\usepackage{cite}
\renewcommand{\baselinestretch}{0.935}

\makeatother

\begin{document}
\title{Collision Diversity SCRAM: Beyond the Sphere-Packing Bound}
\author{\IEEEauthorblockN{Sally Nafie, Joerg Robert, Albert Heuberger}\IEEEauthorblockA{Lehrstuhl für Informationstechnik mit dem Schwerpunkt Kommunikationselektronik
(LIKE)\\
Friedrich-Alexander Universität Erlangen-Nürnberg (FAU),\\
 91058 Erlangen, Germany\\
\{sally.nafie, joerg.robert, albert.heuberger\}@fau.de}%
\noindent\begin{minipage}[t]{1\textwidth}%
This work has been submitted to the IEEE for possible publication.
Copyright may be transferred without notice, after which this version
may no longer be accessible.%
\end{minipage}}
\maketitle
\begin{abstract}
This paper presents a novel scheme dubbed Collision Diversity (CoD)
SCRAM, which is provisioned to meet the challenging requirements of
the future 6G, portrayed in massive connectivity, reliability, and
ultra-low latency. The conventional SCRAM mechanism, which stands
for Slotted Coded Random Access Multiplexing, is a hybrid decoding
scheme, that jointly resolves collisions and decodes the Low Density
Parity Check (LDPC) codewords, in a similar analogy to Belief Propagation
(BP) decoding on a joint three-layer Tanner graph. The CoD SCRAM proposed
herein tends to enhance the performance of SCRAM by adopting an information-theoretic
approach that tends to maximize the attainable Spectral Efficiency.
Besides, due to the analogy between the two-layer Tanner graph of
classical LDPC codes, and the three-layer Tanner graph of SCRAM, the
CoD SCRAM adopts the well-developed tools utilized to design powerful
LDPC codes. Finally, the proposed CoD scheme tends to leverage the
collisions among the users in order to induce diversity. Results show
that the proposed CoD SCRAM scheme surpasses the conventional SCRAM
scheme, which is superior to the state-of-the-art Non-Orthogonal Multiple
Access (NOMA) schemes. Additionally, by leveraging the collisions,
the CoD SCRAM tends to surpass the Sphere-Packing Bound (SPB) at the
respective information block length of the underlying LDPC codes of
the accommodated users. 
\end{abstract}


\section{Introduction }

The future systems are expected to leverage the performance enhancements
which are to be delivered by 6G, in order to promote major features
in diverse fields. 6G is provisioned to play a key role in Healthcare
IoT (HIoT), and Industrial IoT (IIoT), in a multitude of fields such
as logistics, banking, and government sectors \cite{jere2023distributed},
in addition to the deployment of smart sensors that perform self-diagnostics
and autonomous decisions using Artificial Intelligence (AI) \cite{akbar20226g}.
Moreover, 6G is expected to enhance Vehicular IoT (VIoT), by providing
collaboration among connected autonomous vehicles, besides supporting
Brain-Controlled Vehicles (BCV) \cite{nguyen2022towards}. Another
prospect of the future 6G resides in the upbringing of smart cities
in conjunction with the deployment of Digital Twin (DT) \cite{mihai2022digital}.
The realization of these use cases impose constraints on the data
rate, latency, coverage, reliability, connection density, and power
consumption \cite{jiang2021road}. 

Non-Orthogonal Multiple Access (NOMA) \cite{ding2015cooperative}
techniques present themselves as good candidates to support the massive
connectivity constraint. In the literature, various classes of NOMA,
including Power Domain (PD) and Code Domain (CD) NOMA were tackled
\cite{srivastava2021non}. In PD-NOMA \cite{maraqa2020survey}, the
system exploits the disparate power levels in order to differentiate
between the different users. Alternatively, NOMA schemes could be
designed in the code domain\cite{jehan2022comparative,liu2021sparse}.
In the literature, two powerful CD-NOMA schemes stand out; Pattern
Division Multiple Access (PDMA) \cite{chen2016pattern} and Sparse
Code Multiple Access (SCMA) \cite{nikopour2013sparse}. These techniques
fall under the umbrella of codebook-based NOMA \cite{de2019random},
where the analogy between Low Density Spreading (LDS) \cite{hoshyar2008novel}
and factor graph representation of codes on graphs \cite{loeliger2004introduction}
such as LDPC codes, was exploited to facilitate the Multiuser Detection
(MUD). 

In addition, sequence-based NOMA schemes, that rely on spreading sequences,
were proposed in an attempt to optimize the tradeoff between spectral
efficiency, receiver performance, and MUD complexity \cite{sosnin2017non,dai2018survey}.
Besides, Interleaved-based NOMA exploits the idea of uniquely designed
interleavers as a candidate option to differentiate between the users
\cite{dai2018survey}. Interleave Division Multiple Access (IDMA)
\cite{ping2006interleave} is one of the viable Interleave-based NOMA
schemes. 

In order to provide reliability, most of the proposed NOMA schemes
rely on packet repetition or a variant thereof, which does not fully
utilize the degrees of freedom. Moreover, the key terminology of the
the proposed NOMA schemes depends on thoroughly coordinating the resource
allocation among the users. Consequently, the proposed techniques
perform poorly in terms of scalability. It is worth mentioning that
in order to meet the low latency requirement, some of the aforementioned
NOMA schemes propose a grant-free variant of their respective approach.
However, a grant-free approach would violate the key essence of these
schemes, which rely mainly on attempting to reduce the cross-correlation
among the collided packets, for them to be able to resolve collisions. 

The NOMA schemes presented in the literature adopt a sequential decoding
scheme, portrayed in a Multiuser Detector (MUD), followed by a bank
of Forward Error Correction (FEC) decoders. The MUD detector could
vary between Successive Interference Cancellation \cite{iswarya2021survey},
Belief Propagation \cite{tan2006belief}, or Expectation Propagation
Algorithm \cite{meng2017low}. The output of the MUD detector is then
fed to the bank of FEC decoders. Turbo-like Decoders (Iterative Detection
and Decoding (IDD)) \cite{ren2016advanced}, involve feeding the output
of the bank of the FEC decoders back to the input of the MUD detector.
This could be regarded as an iterative sequential decoding scheme. 

Unlike the sequential decoders presented in the literature, a joint
decoder dubbed SCRAM, which stands for Slotted Coded Random Access
Multiplexing, is proposed in \cite{nafie2018scram}. SCRAM combines
the reliability of LDPC codes \cite{gallager1962low}, with the low-latency
privilege of Slotted ALOHA (SA) \cite{munari2013throughput}. In a
similar analogy to Belief Propagation (BP) \cite{mackay1997near}
decoding of LDPC codes on a two-layer Tanner graph \cite{tanner1981recursive},
the joint SCRAM decoder concurrently resolves collisions and decodes
the LDPC codewords, on a joint three-layer Tanner graph. 

In this paper, a novel approach dubbed Collision Diversity SCRAM (CoD
SCRAM), that tends to further enhance the performance of SCRAM, is
proposed. The essence of the CoD SCRAM scheme is threefold; First,
the CoD scheme follows an information theoretic perspective in an
attempt to maximize the attainable spectral efficiency. Secondly,
due to the analogy between the two-layer Tanner graph of classical
LDPC codes, and the three-layer Tanner graph of SCRAM, the proposed
CoD SCRAM is designed such that it alleviates the detrimental factors
that deteriorate the performance of classical LDPC codes. Finally,
the proposed CoD SCRAM scheme leverages the collisions among the users
in order to gain diversity.

The rest of this paper is organized as follows. Section II tackles
the SCRAM system model, the Tanner Graph representation, and joint
BP decoding. In Section III, the theoretical capacity bounds of the
overloaded shared channel are derived. The design aspects of the classical
LDPC Tanner graph are tackled in Section IV. Next, Section V presents
the novel Collision Diversity SCRAM scheme. Then, Section VI numerically
assesses the graphical structure of the proposed CoD SCRAM, revisiting
the design aspects tackled in Section IV. The performance of the proposed
CoD SCRAM is assessed in Section VI via Monte Carlo simulations. Finally,
Section VII summarizes the paper and concludes the results.

\section{SCRAM Preliminaries}

\subsection{System Model}

The SCRAM system incorporates $N_{u}$ users sharing the SA wireless
medium. Each user $U_{n_{u}},\;\text{\ensuremath{\forall}}n_{u}=1,\ldots,N_{u}$,
has an information packet, $\mathbf{b^{\left(\mathit{n_{u}}\right)}}$,
of length $k_{n_{u}}$ bits. This packet is encoded with an LDPC encoder,
of code rate $r_{n_{u}}=k_{n_{u}}/n_{n_{u}}$, producing an output
codeword, $\mathbf{c^{\left(\mathit{n_{u}}\right)}}$, of length $n_{n_{u}}$
bits. Without loss of generality, the encoded bits are mapped to BPSK
modulated symbols. The vector of modulated symbols of user $U_{n_{u}}$,
is given by $\mathbf{x^{\left(\mathit{n_{u}}\right)}}$. The BPSK
symbols are then transmitted using OFDM. Incorporating OFDM is twofold:
to provide slot synchronization due to the gridded analogy between
SA and OFDM subcarriers, and to translate the dispersive multipath
fading channel to multiple flat fading subchannels. Prior to transmission,
each user, $U_{n_{u}}$, randomly and independently chooses $n_{n_{u}}$
SA slots (OFDM subcarriers) to transmit its modulated encoded codeword.
Let $N_{s}$ denote the total number of available slots per SA frame.
In this case, the channel load -- defined as the number of useful
information bits per SA slot -- is given by $D=\sum_{n_{u}=1}^{N_{u}}k_{n_{u}}/N_{s}$.

The transmitted symbols are affected by multipath fading, and perturbed
by Additive White Gaussian Noise (AWGN). Due to the random selection
of SA, some slots are collision free, while others suffer from collisions.
The received signal at a contended slot corresponds to the superposition
of the collided faded symbols added to the AWGN.

At the receiver side, a joint decoding of the SA contended LDPC codewords
is performed iteratively. The essence of SCRAM lies in the joint decoding
of both SA and LDPC. Unlike the sequential decoders presented in the
literature, the SCRAM mechanism incorporates a parallel three-layer
Tanner graph that allows for the joint decoding. The idea is inspired
by BP decoding of graph-based codes such as LDPC codes.

\subsection{Three-Layer Tanner Graph Representation}

For a SCRAM system that accommodates $N_{u}$ users, the corresponding
joint three-layer Tanner graph of the SCRAM decoder, comprises a set
of variable nodes representing the transmitted symbols from all the
users. These variable nodes are bounded by two layers of check nodes.
The first check nodes layer corresponds to the slots of the shared
SA medium. The other layer is driven from the conventional parity
check nodes in the BP decoding of classical LDPC codes. Let $N_{v}$
denote the number of variable nodes. Each transmitted symbol is represented
by a variable node. Thus, $N_{v}=\sum_{n_{u}=1}^{^{N_{u}}}n_{n_{u}}$,
where $n_{n_{u}}$ represents the number of transmitted symbols of
user $U_{n_{u}}$ per SA frame. The number of SA check nodes, denoted
by $N_{s}$, is allocated according to the available radio resources.
The total number of LDPC check nodes is given by $N_{l}=\sum_{n_{u}=1}^{^{N_{u}}}m_{n_{u}}$,
where $m_{n_{u}}\geq n_{n_{u}}-k_{n_{u}}$, denotes the number of
LDPC parity check equations of user $U_{n_{u}}$.

The connections between the variable nodes and the SA check nodes
rely on the random selection of the transmission slots of each user.
A contended SA check node can be connected to more than one variable
node. Meanwhile, the connections of the variable nodes to the LDPC
check nodes are fully determined by the deployed LDPC encoder at each
user's transmit terminal. This is exactly the same as the conventional
representation of LDPC codes on Tanner graphs.

\begin{figure}[b]
\begin{centering}
\includegraphics[width=1\columnwidth]{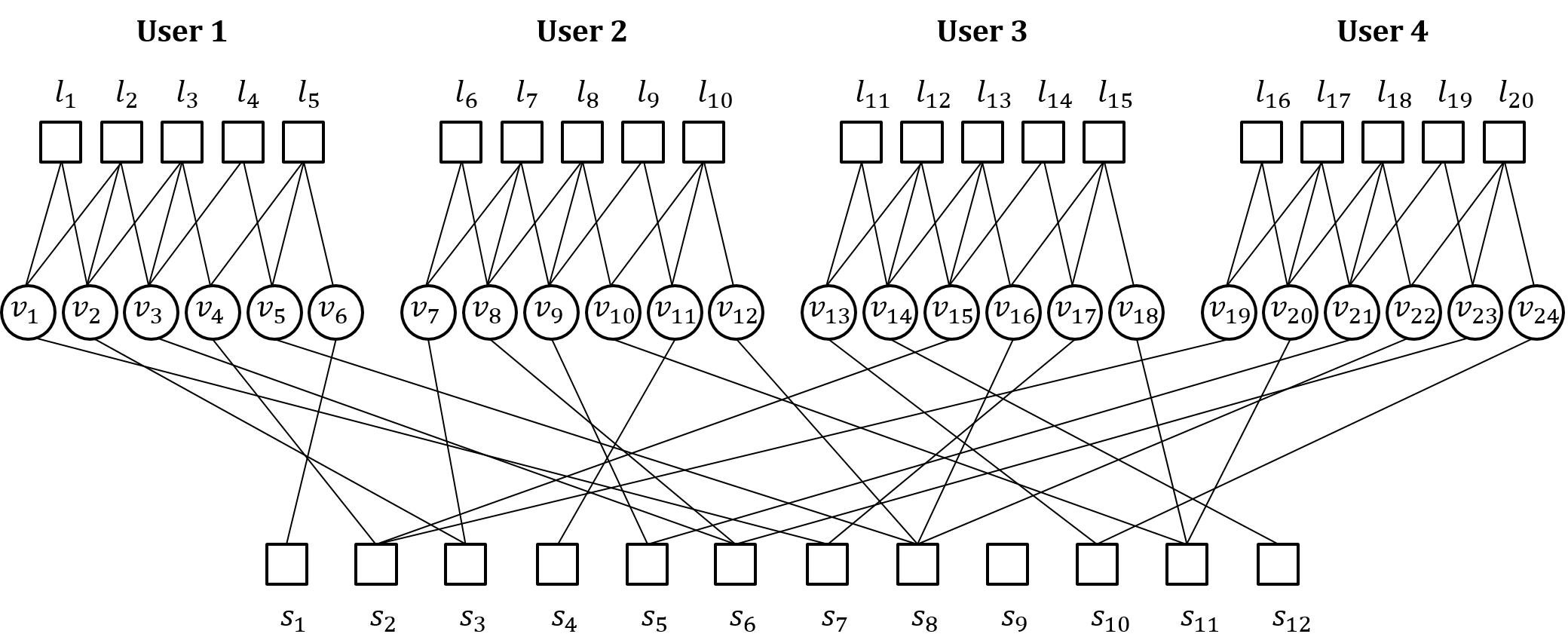}
\par\end{centering}
\caption{Three-Layer Tanner graph example of SCRAM system with $N_{u}=4$ users,
each tranmitting $n_{n_{u}}=6$ LDPC encoded symbols over a system
with $N_{s}=12$ SA slots}
\label{Tanner Example}

\end{figure}

For illustration, consider the three-layer Tanner graph of the SCRAM
model shown in Figure \ref{Tanner Example}. The model incorporates
$N_{u}=4$ users, each transmitting $n_{n_{u}}=6$ modulated symbols.
For simplicity, it is assumed that the four users adopt identical
LDPC encoders with $n_{n_{u}}=6$ variable nodes, and $m_{n_{u}}=5$
LDPC check nodes, per user. The three-layer Tanner graph of such a
model consists of $N_{v}=24$ variable nodes, $N_{l}=20$ LDPC check
nodes. Moreover, the graph has a set of $N_{s}=12$ SA check nodes,
that correspond to 12 allocated frequency subcarriers. Each user blindly
selects six subcarriers to transmit its six modulated symbols, without
prior knowledge about the selected slots of the other three users.

\subsection{Iterative Joint SCRAM Decoding}

The objective of the SCRAM decoder is to resolve the contended packets
through a hybrid iterative decoding scheme, that is jointly carried
out on both LDPC and SA. Inspired by Belief Propagation (BP) decoding
\cite{mackay1997near} of classical LDPC codes, each decoding iteration
involves two message passing steps; check nodes to variable nodes,
and variable nodes to check nodes. In every iteration, the variable
nodes concurrently communicate with both the SA and the LDPC check
nodes, on the three-layer Tanner graph that comprises $N_{v}$ variable
nodes, $N_{s}$ SA check nodes, and $N_{l}$ LDPC check nodes.

\subsubsection{Check Nodes to Variable Nodes}

In the first half of the iteration, the SA and LDPC check nodes send
their provisioned Log Likelihood Ratios (LLRs) to their corresponding
variable nodes. The LLR transmitted from one check node to its corresponding
variable node reflects the reliability of the transmitted value represented
by this variable node.

\paragraph{SA Check Nodes to Variable Nodes}

SA check node $s_{n_{s}}$ sends an LLR to variable node $v_{n_{v}}$
if they exhibit a connection in the associated three-layer Tanner
graph. The LLR value $S_{n_{s},n_{v}}$ transmitted from SA check
node $s_{n_{s}}$ to variable node $v_{n_{v}}$ is given by
\begin{equation}
S_{n_{s},n_{v}}=\ln\left[\frac{\ensuremath{\sum\limits _{\acute{m}=1}^{M_{n_{s}}^{\left(\pm\right)}}\left[\mathrm{P}(y=y_{n_{s}}|\check{\mathbf{x}}_{\acute{m}}^{\left(n_{v},+1\right)})\cdot\mathrm{P}\left(\check{\mathbf{x}}_{\acute{m}}^{\left(n_{v},+1\right)}\right)\right]}}{\sum\limits _{\acute{m}=1}^{M_{n_{s}}^{\left(\pm\right)}}\left[\mathrm{P}(y=y_{n_{s}}|\check{\mathbf{x}}_{\acute{m}}^{\left(n_{v},-1\right)})\cdot\mathrm{P}\left(\check{\mathbf{x}}_{\acute{m}}^{\left(n_{v},-1\right)}\right)\right]}\right],\label{eq:21}
\end{equation}

\noindent where for $n_{s}=1,\cdots,N_{s}$, $y_{n_{s}}$ represents
the channel received signal at SA slot $s_{n_{s}}$. Let $\mathcal{X}_{n_{s}}$,
bet the set of vectors of all the possible combinations of the collided
symbols at SA check node $s_{n_{s}}$. Moreover, let $\mathcal{X}_{n_{s},+1}^{\left(n_{v}\right)}$
and $\mathcal{X}_{n_{s},-1}^{\left(n_{v}\right)}$ be a subset of
$\mathcal{X}_{n_{s}}$ that represents the set of possible vectors
of all the collided symbols at SA check node $s_{n_{s}}$, that assume
that the symbol represented by variable node $v_{n_{v}}$ is $+1$
and $-1$, respectively. The cardinality of both $\mathcal{X}_{n_{s},+1}^{\left(n_{v}\right)}$
and $\mathcal{X}_{n_{s},-1}^{\left(n_{v}\right)}$, denoted by $M_{n_{s}}^{\left(\pm\right)}$,
amounts to half the cardinality of $\mathcal{X}_{n_{s}}$. For $\acute{m}=1,\cdots,M_{n_{s}}^{\left(\pm\right)}$,
let $\mathbf{\mathbf{\check{x}_{\mathit{\acute{m}}}^{\left(\mathit{n_{v},\mathrm{+1}}\right)}}}$
and $\mathbf{\mathbf{\check{x}_{\mathit{\acute{m}}}^{\left(\mathit{n_{v},\mathrm{-1}}\right)}}}$
be a vector that corresponds to one of the $M_{n_{s}}^{\left(\pm\right)}$
combination vectors of $\mathcal{X}_{n_{s},+1}^{\left(n_{v}\right)}$
and $\mathcal{X}_{n_{s},-1}^{\left(n_{v}\right)}$, respectively.
Finally, $\mathrm{P}(\mathbf{\mathbf{\check{x}_{\acute{\mathit{m}}}^{\left(\mathit{n_{v},\mathrm{+1}}\right)}}})$
and $\mathrm{P}(\mathbf{\mathbf{\check{x}_{\acute{\mathit{m}}}^{\left(\mathit{n_{v},\mathrm{-1}}\right)}}})$
denote the apriori probability of the vectors $\mathbf{\mathbf{\check{x}_{\mathit{\acute{m}}}^{\left(\mathit{n_{v},\textrm{+1}}\right)}}}$
and $\mathbf{\mathbf{\check{x}_{\mathit{\acute{m}}}^{\left(\mathit{n_{v},\textrm{-1}}\right)}}}$,
respectively.

\noindent For $\acute{m}=1,\cdots,M_{n_{s}}^{\left(\pm\right)}$,
the conditional probabilities of the vector $\check{\mathbf{x}}_{\mathit{\acute{m}}}^{\left(\mathit{n_{v}},+1\right)}\subset\mathcal{X}_{n_{s},+1}^{\left(n_{v}\right)}$,
and the vector $\check{\mathbf{x}}_{\mathit{\acute{m}}}^{\left(\mathit{n_{v}},-1\right)}\subset\mathcal{X}_{n_{s},-1}^{\left(n_{v}\right)}$,
are respectively given by

\begin{equation} \resizebox{1.0\hsize}{!}{$\mathrm{P}\left(y=y_{n_{s}}|\mathbf{\check{x}_{\acute{\mathit{m}}}^{\left(\mathit{n_{v}}\mathrm{,\pm1}\right)}}\right)=\frac{1}{\pi\sigma^{2}}\exp\left[\frac{\left|y_{n_{s}}-\ensuremath{\sum\limits _{\acute{d}=1}^{d_{s_{n_{s}}}}h_{s_{n_{s}},\acute{d}}\;}\check{x}_{\acute{m},\acute{d}}^{\left(n_{v},\pm1\right)}\right|^{2}}{\sigma^{2}}\right],$} \end{equation}where
$\check{x}_{\acute{m},\acute{d}}^{\left(n_{v},+1\right)}$and $\check{x}_{\acute{m},\acute{d}}^{\left(n_{v},-1\right)}$
represent a possible value for the modulated symbol represented by
the $\acute{d}^{\textrm{th}}$ variable node that collides at SA check
node $s_{n_{s}}$, in the vector $\check{\mathbf{x}}_{\acute{\mathit{m}}}^{\left(\mathit{n_{v}}\mathrm{,+1}\right)}$and
$\mathbf{\check{x}_{\mathit{\acute{m}}}^{\left(\mathit{n_{v}}\mathrm{,-1}\right)}}$,
respectively. Moreover, $h_{s_{n_{s}},\acute{d}}$ denotes the estimated
fading coefficient of the channel between SA check node $s_{n_{s}}$
and the $\acute{d}^{\textrm{th}}$ variable node that collides at
it. Finally, $\sigma^{2}$ is the noise variance of the complex AWGN
channel.

At the first iteration, for $\acute{m}=1,\cdots,M_{n_{s}}^{\left(\pm\right)}$,
the apriori probability of the vector $\check{\mathbf{x}}_{\mathit{\acute{m}}}^{\left(\mathit{n_{v}},+1\right)}\subset\mathcal{X}_{n_{s},+1}^{\left(n_{v}\right)}$,
or the vector $\check{\mathbf{x}}_{\mathit{\acute{m}}}^{\left(\mathit{n_{v}},-1\right)}\subset\mathcal{X}_{n_{s},-1}^{\left(n_{v}\right)}$,
is given by 
\begin{equation}
\mathrm{P}(\mathbf{\check{x}_{\acute{\mathit{m}}}^{\left(\mathit{n_{v}}\mathrm{,+1}\right)}})=\mathrm{P}(\mathbf{\check{x}_{\acute{\mathit{m}}}^{\left(\mathit{n_{v}}\mathrm{,-1}\right)}})=\frac{1}{M_{n_{s}}^{\left(\pm\right)}}.
\end{equation}

Starting from the second iteration, for $\acute{m}=1,\cdots,M_{n_{s}}^{\left(\pm\right)}$,
the apriori probabilities of the vector $\check{\mathbf{x}}_{\mathit{\acute{m}}}^{\left(\mathit{n_{v}},+1\right)}$,
and the vector $\check{\mathbf{x}}_{\mathit{\acute{m}}}^{\left(\mathit{n_{v}},-1\right)}$,
are respectively given by 

\begin{equation}
\mathrm{P}(\mathbf{\check{x}_{\mathit{\acute{m}}}^{\left(\mathit{n_{v}}\mathrm{,\pm1}\right)}})=\ensuremath{\prod_{\substack{\acute{d}=1\\
v_{n_{v}}\neq v_{s_{n_{s}},\acute{d}}
}
}^{d_{s_{n_{s}}}}\mathrm{P}\left(\check{x}_{\acute{m},\acute{d}}^{\left(n_{v},\pm1\right)}\right)},\label{eq:25}
\end{equation}

\noindent where $\mathrm{P}\left(\check{x}_{\acute{m},\acute{d}}^{\left(n_{v},\pm1\right)}\right)$
denotes the apriori probability of the $\acute{d}^{\textrm{th}}$
symbol, in the vector $\check{\mathbf{x}}_{\acute{m}}^{\left(n_{v},+1\right)}$
and $\check{\mathbf{x}}_{\acute{m}}^{\left(n_{v},-1\right)}$, respectively.

Let $V_{n_{s},\acute{n_{v}}}^{\left(S\right)}$ denote the LLR that
SA check node $s_{n_{s}}$, received in the previous iteration, from
its corresponding variable node $v_{\acute{n_{v}}}$, such that, $v_{\acute{n_{v}}}=v_{s_{n_{s}},\acute{d}}$,
denotes the $\acute{d}^{\textrm{th}}$ variable node in the set of
associated variable nodes of SA check node $s_{n_{s}}$. Upon receiving
these LLRs, the SA check node, $s_{n_{s}}$, updates the apriori probability
of each symbol that corresponds to one of its associated variable
nodes. The updated apriori probability of $\check{x}_{\acute{m},\acute{d}}^{\left(n_{v},+1\right)}\in\check{\mathbf{x}}_{\mathit{\acute{m}}}^{\left(\mathit{n_{v}}\mathrm{,+1}\right)}$,
and $\check{x}_{\acute{m},\acute{d}}^{\left(n_{v},-1\right)}\in\check{\mathbf{x}}_{\mathit{\acute{m}}}^{\left(\mathit{n_{v}}\mathrm{,-1}\right)}$,
are respectively calculated as follows

\noindent 
\begin{equation}
\mathrm{P}\left(\check{x}_{\acute{m},\acute{d}}^{\left(n_{v},\pm1\right)}\right)=\begin{cases}
\frac{1}{1+\exp\left[-V_{n_{s},\acute{n_{v}}}^{\left(S\right)}\right]} & \text{{for\ }}\check{x}_{\acute{m},\acute{d}}^{\left(n_{v},\pm1\right)}=+1\\
\frac{-V_{n_{s},\acute{n_{v}}}}{1+\exp\left[-V_{n_{s},\acute{n_{v}}}^{\left(S\right)}\right]} & \text{{for\ }}\check{x}_{\acute{m},\acute{d}}^{\left(n_{v},\pm1\right)}=-1
\end{cases}.\label{eq:27}
\end{equation}

\paragraph{LDPC Check Nodes to Variable Nodes}

The LLR transmitted from the LDPC check nodes to their corresponding
variable nodes are calculated in the same way as the classical LDPC
BP algorithm. The extrinsic information transmitted from an LDPC check
node to one of its associated variable nodes represents the updated
LLR of the probability of the corresponding symbol represented by
the variable node, that causes the LDPC check node to be satisfied.

An LDPC check node, $l_{n_{l}}$, sends an LLR to a variable node,
$v_{n_{v}}$, if and only if $v_{n_{v}}$ belongs to the set of associated
variable nodes of LDPC check node $l_{n_{l}}$. The LLR value, $L_{n_{l},n_{v}}$,
transmitted from LDPC check node $l_{n_{l}}$, to variable node $v_{n_{v}}$,
is given by
\begin{equation}
L_{n_{l},n_{v}}=-2\tanh^{-1}\left(\prod_{\substack{v_{\acute{n_{v}}}\epsilon\mathbf{v}_{l_{n_{l}}}\\
v_{\acute{n_{v}}}\neq v_{n_{v}}
}
}\tanh\left(-\frac{V_{n_{l},\acute{n_{v}}}^{\left(L\right)}}{2}\right)\right),
\end{equation}
where $V_{n_{l},\acute{n_{v}}}^{\left(L\right)}$ represents the LLR
transmitted from variable node $v_{\acute{n_{v}}}$, to LDPC check
node $l_{n_{l}}$, in the previous iteration.

\subsubsection{Variable Nodes to Check Nodes}

Upon receiving the LLRs from the SA and the LDPC check nodes, the
variable nodes calculate new information to be transmitted to their
corresponding check nodes, both SA and LDPC, in the next half of the
iteration. This information represents the LLR of their currently
held symbols, based on the information they gathered from their corresponding
SA and LDPC check nodes. The key essence of the joint SCRAM decoding
algorithm is the simultaneous transmission of the variable nodes to
both SA and LDPC check nodes.

Variable node, $v_{n_{v}}$, would send an LLR to SA check node $s_{n_{s}}$,
and to to LDPC check node $l_{n_{l}}$, if and only if SA check node
$s_{n_{s}}$, and LDPC check node $l_{n_{l}}$, belong to the set
of associated SA check nodes, and LDPC check nodes, respectively,
of variable node, $v_{n_{v}}$. Let $V_{n_{s},n_{v}}^{\left(S\right)}$
and $V_{n_{l},n_{v}}^{\left(L\right)}$ represent the information
transmitted from variable node, $v_{n_{v}}$ to SA check node $s_{n_{s}}$,
and LDPC check node $l_{n_{l}}$, respectively. These LLRs are respectively
given by 
\begin{equation}
V_{n_{s},n_{v}}^{\left(S\right)}=\sum_{\substack{s_{\acute{n_{s}}}\in\boldsymbol{\mathbf{s}}_{v_{n_{v}}}\\
s_{\acute{n_{s}}}\neq s_{n_{s}}
}
}S_{\acute{n_{s}},n_{v}}+\ensuremath{\sum\limits _{l_{\acute{n_{l}}}\in\mathbf{\mathbf{l}}_{v_{n_{v}}}}L_{\acute{n_{l}},n_{v}}},\label{eq:30}
\end{equation}

\begin{equation}
V_{n_{l},n_{v}}^{\left(L\right)}=\ensuremath{\sum\limits _{s_{\acute{n_{s}}}\in\mathbf{\boldsymbol{s}}_{v_{n_{v}}}}S_{\acute{n_{s}},n_{v}}}+\sum_{\substack{l_{\acute{n_{l}}}\in\mathbf{\mathbf{l}}_{v_{n_{v}}}\\
l_{\acute{n_{l}}}\neq l_{n_{l}}
}
}L_{\acute{n_{l}},n_{v}},\label{eq:31}
\end{equation}
where $S_{\acute{n_{s}},n_{v}}$ and $L_{\acute{n_{l}},n_{v}}$ represent
the LLR, that variable node, $v_{n_{v}}$, has received in the first
half of the iteration from SA check node, $s_{\acute{n_{s}}}$, and
LDPC check node, $l_{\acute{n_{l}}}$, respectively. 

\section{A Glimpse on Information Theory and Theoretical Capacity Bounds}

\subsection{Derivation of Collision-based Channel Capacity}

\subsubsection{Random Access}

In this section, the multiuser channel capacity on the Random Access
(RA) channel is derived. The adopted RA model is the Slotted ALOHA
(SA) \cite{munari2013throughput}. For SA, the channel is equally
divided into $N_{s}$ slots of equal width. These slots could either
be time slots or frequency subcarriers.

Let $N_{u}$ be the number of users accessing an SA frame that comprises
$N_{s}$ slots. Despite the fact the multiple users could collide
at the same slot, the collision of two or more symbols of the same
user on one SA slot is not possible \cite{casini2007contention}.
This means that the number of collisions on one SA slot ranges from
0 to $N_{u}$. For $n_{s}=1,\ldots,N_{s}$, the slot degree, $d_{s_{n_{s}}}$,
such that $0\leq d_{s_{n_{s}}}\leq N_{u}$, is defined as the number
of collided symbols (or users) at SA slot, $s_{n_{s}}$. For this
model, the node degrees follow a binomial distribution \cite{altham1978two}.
This means that the probability of SA slot $s_{n_{s}}$, having a
certain degree, $0\leq d_{s_{n_{s}}}\leq N_{u}$, is given by
\begin{equation}
\textrm{P}\left(d_{s_{n_{s}}}\right)={N_{u} \choose d_{s_{n_{s}}}}p^{d_{s_{n_{s}}}}\left(1-p\right)^{N_{u}-d_{s_{n_{s}}}},\;0\leq d_{s_{n_{s}}}\leq N_{u},\label{eq: Binomial distribution}
\end{equation}
where $p$ denotes the probability that SA slot, $s_{n_{s}}$, is
chosen by user, $U_{n_{u}}$. 

\begin{figure}[tbh]
\centering{}\includegraphics[width=1\columnwidth]{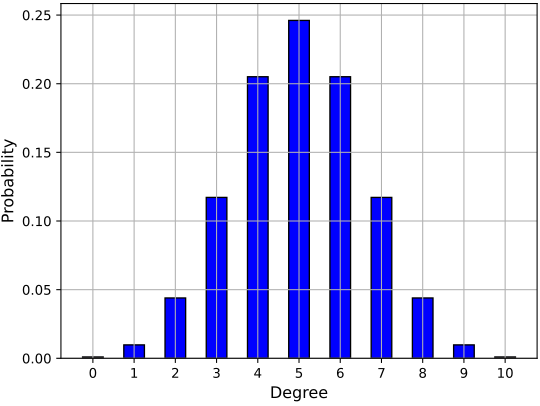}\caption{Simulated degree distribution of an SA system with $N_{s}=8064$ slots,
incorporating $N_{u}=10$ users, each transmitting $n_{n_{u}}=4320$
bits\label{fig:Simulated-degree-distribution}}
\end{figure}

Because of the random access, all $N_{s}$ SA slots are equally-likely
chosen. For $n_{u}=1,\ldots,N_{u}$, if user $U_{n_{u}}$ attempts
to transmit $n_{n_{u}}$ symbols over an SA frame of $N_{s}$ slots,
the probability, $p$, of an arbitrary SA slot being chosen by user
$U_{n_{u}}$, to transmit one of its $n_{n_{u}}$ symbols is given
by $p=\frac{n_{n_{u}}}{N_{s}}$. Similarly, because the users randomly
choose the slots, (\ref{eq: Binomial distribution}) applies for all
the $N_{s}$ slots. Figure \ref{fig:Simulated-degree-distribution}
depicts the simulated degree distribution of an SA system with $N_{s}=8640$
slots, incorporating $N_{u}=10$ users, where each user transmits
$n_{n_{u}}=4320$ bits (symbols). As shown in the figure, the degree
distribution ranges from zero to ten collisions per slot. The simulated
probabilities coincide with the binomial distribution of (\ref{eq: Binomial distribution}),
computed at $p=0.5$.

The key to deriving the capacity of this system, is to first derive
the capacity of one SA slot, then extend the derivation to $N_{s}$
slots. For one SA slot, the achievable sum rate, $C_{s_{n_{s}}}$,
at SA slot $s_{n_{s}}$, is given by

\begin{equation}
C_{s_{n_{s}}}=\ensuremath{\sum\limits _{d_{s_{n_{s}}}=0}^{N_{u}}\textrm{P}\left(d_{s_{n_{s}}}\right)C_{d_{s_{n_{s}}}}\;\textrm{bits/s/SAslot.}}
\end{equation}

Assuming each SA slot, $s_{n_{s}}$, has a bandwidth of $W_{n_{s}}=\frac{W}{N_{s}}$,
the sum rate, $C_{d_{s_{n_{s}}}}$, corresponds to the classical single-user
capacity with bandwidth $W_{n_{s}}$, and received power, $d_{s_{n_{s}}}\frac{P}{N_{u}n_{n_{u}}}$.
Consequently, the overall sum rate, $C_{s_{n_{s}}}$, at SA slot $s_{n_{s}}$,
in compliance with \cite{proakis2008digital}, can be written as

\begin{equation} \resizebox{1.0\hsize}{!}{$C_{s_{n_{s}}}=\ensuremath{\sum\limits _{d_{s_{n_{s}}}=0}^{N_{u}}\textrm{P}\left(d_{s_{n_{s}}}\right)\frac{W}{N_{s}}\log_{2}\left(1+\frac{d_{s_{n_{s}}}\frac{P}{N_{u}n_{n_{u}}}}{N_{0}\frac{W}{N_{s}}}\right)\;\textrm{bits/s/SAslot.}}$} \end{equation}

This means that the overall sum rate, $C_{SA}$, of the SA system
with $N_{s}$ slots is given by

\begin{equation} \resizebox{1.0\hsize}{!}{$C_{SA}=N_{s}C_{s_{n_{s}}}=\ensuremath{\sum\limits _{d_{s_{n_{s}}}=0}^{N_{u}}\textrm{P}\left(d_{s_{n_{s}}}\right)W\log_{2}\left(1+\frac{d_{s_{n_{s}}}\frac{P}{N_{u}n_{n_{u}}}}{N_{0}\frac{W}{N_{s}}}\right)\;\textrm{bits/s.}}$} \end{equation}

The total transmit power per user, $\frac{P}{N_{u}}$, is equivalent
to $E_{b}R_{b}$, where $E_{b}$ and $R_{b}$ denote the energy per
bit, and the transmission data rate per user, respectively. Thus,
the overall sum rate can be further simplified to

\begin{equation}
C_{SA}=\ensuremath{\sum\limits _{d_{s_{n_{s}}}=0}^{N_{u}}\textrm{P}\left(d_{s_{n_{s}}}\right)W\log_{2}\left(1+\frac{E_{b}}{N_{0}}\frac{N_{s}d_{s_{n_{s}}}}{n_{n_{u}}}\frac{R_{b}}{W}\right)\;\textrm{bits/s.}}
\end{equation}

Moreover, the overall spectral efficiency, $\eta_{SA}$, defined as
the ratio between the overall sum rate, $C_{SA}$, and the total bandwidth,
$W$, is given by

\begin{equation}
\eta_{SA}=\ensuremath{\sum\limits _{d_{s_{n_{s}}}=0}^{N_{u}}\textrm{P}\left(d_{s_{n_{s}}}\right)\log_{2}\left(1+\frac{E_{b}}{N_{0}}\frac{N_{s}d_{s_{n_{s}}}}{n_{n_{u}}}\frac{R_{b}}{W}\right)\;\textrm{bits/s/Hz.}}
\end{equation}

Finally, substituting for $\frac{R_{b}}{W}$ by $\frac{\eta_{SA}}{N_{u}}$,
the overall spectral efficiency can be written as

\begin{equation}
\eta_{SA}=\ensuremath{\sum\limits _{d_{s_{n_{s}}}=0}^{N_{u}}\textrm{P}\left(d_{s_{n_{s}}}\right)\log_{2}\left(1+\frac{E_{b}}{N_{0}}\frac{N_{s}d_{s_{n_{s}}}}{N_{u}n_{n_{u}}}\eta_{SA}\right)\;\textrm{bits/s/Hz.}}\label{eq: capacity_RA}
\end{equation}

\subsubsection{Uniform Access}

Similar to the random access channel, a multiuser uniform access channel
allows for overloading. However, the uniform access imposes subtle
coordination among the users, such that the number of collisions per
each slot is the same.

Let the uniform multiple access channel be divided into $N_{s}$ slots
of equal width in the frequency domain (Mapping to time domain is
straightforward). Assuming that $N_{u}$ users, with $n_{n_{u}}$
symbols each, attempt to share the channel, the uniform degree, $d_{s_{n_{s}}}$,
of each slot, $s_{n_{s}}$, is given by

\begin{equation}
d_{s_{n_{s}}}=\frac{N_{u}n_{n_{u}}}{N_{s}},
\end{equation}
assuming that $\textrm{mod}\left(N_{u}n_{n_{u}},N_{s}\right)=0$.
Mapping to the case where $\textrm{mod}\left(N_{u}n_{n_{u}},N_{s}\right)\neq0$
is straightforward. In that case, the overall sum rate, $C_{s_{n_{s}}}$,
at slot $s_{n_{s}}$, can be written as

\[
C_{s_{n_{s}}}=\ensuremath{\frac{W}{N_{s}}\log_{2}\left(1+\frac{d_{s_{n_{s}}}\frac{P}{N_{u}n_{n_{u}}}}{N_{0}\frac{W}{N_{s}}}\right)}
\]

\[
=\frac{W}{N_{s}}\log_{2}\left(1+\frac{\frac{P}{N_{s}}}{N_{0}\frac{W}{N_{s}}}\right)
\]

\begin{equation}
=\ensuremath{\frac{W}{N_{s}}\log_{2}\left(1+\frac{P}{N_{0}W}\right)\;\textrm{bits/s/slot}.}
\end{equation}

Consequently, the overall sum rate of the uniform-access channel with
$N_{s}$ slots is given by

\[
C_{uniform}=\ensuremath{W\log_{2}\left(1+\frac{P}{N_{0}W}\right)}
\]

\begin{equation}
=\ensuremath{W\log_{2}\left(1+\frac{E_{b}}{N_{0}}N_{u}\frac{R_{b}}{W}\right)\;\textrm{bits/s.}}
\end{equation}

Thus, the overall spectral efficiency, $\eta_{uniform}$, can be written
as

\begin{equation}
\eta_{uniform}=\ensuremath{\log_{2}\left(1+\frac{E_{b}}{N_{0}}\eta_{uniform}\right)\;\textrm{bits/s/Hz.}}\label{eq:capacity uniform}
\end{equation}

Finally, the nominal SNR per bit, required to meet a specific target
spectral efficiency, can be calculated as

\begin{equation}
\frac{E_{b}}{N_{0}}=\frac{2^{\eta_{uniform}}-1}{\eta_{uniform}}.
\end{equation}

\subsection{Trade-off between Random Access and Uniform Access}

As discussed in the previous section, a shared random access channel,
that accommodates $N_{u}$ users, is attributed to a binomial distribution
of its channel slot degree. This means that the random access channel
attains a non-zero probability of idle slots. A uniform access channel,
on the other hand, although allows for collisions, maintains a uniform
degree distribution to all channel slots. This means that unlike random
access, unless fully idle, a uniform access channel has strictly zero
probability of possessing idle slots.

\begin{figure}[b]
\begin{centering}
\includegraphics[width=1\columnwidth]{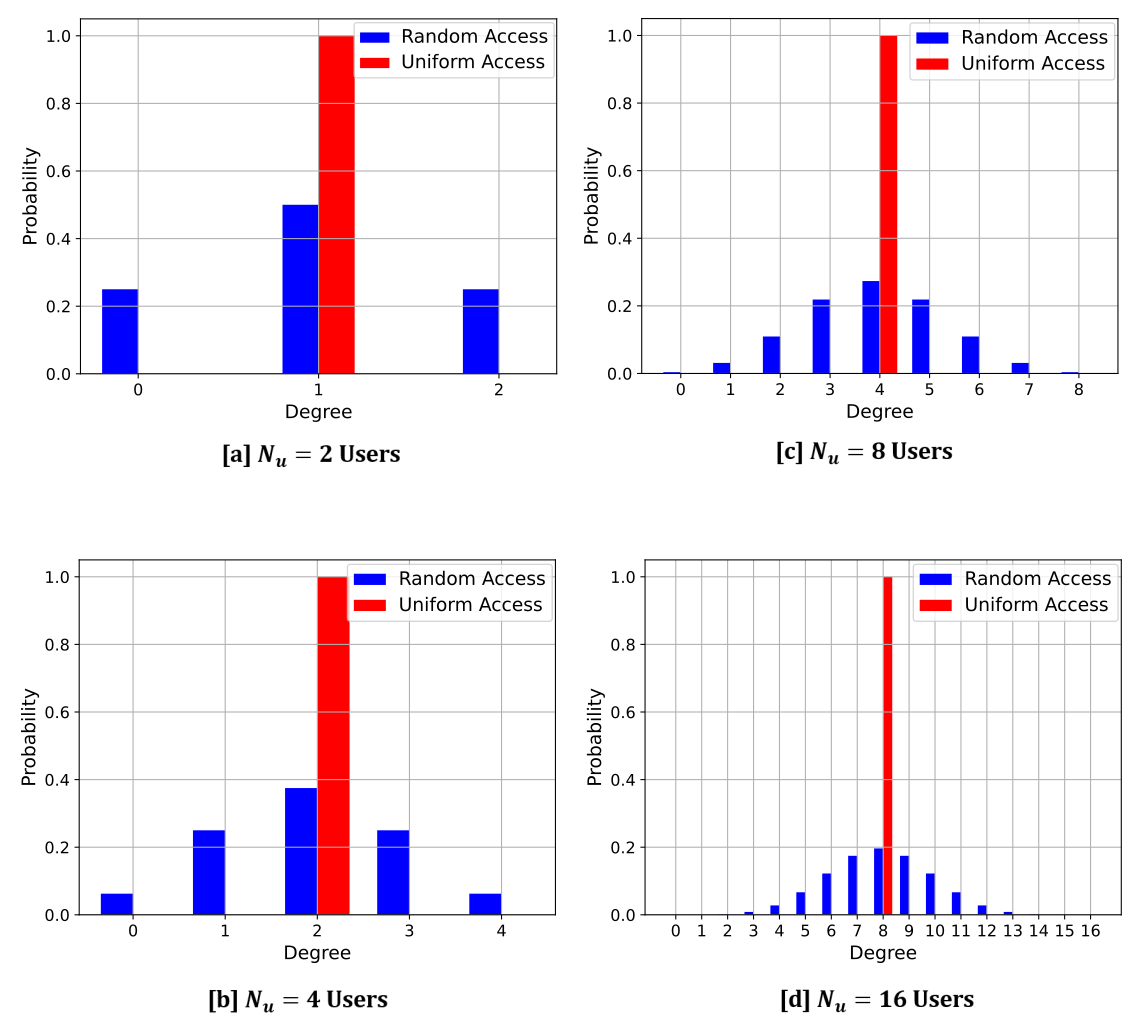}\caption{Node degree distribution comparison of Random and Uniform Channel
Access}
\label{fig:Binomial Vs Uniform Distribution}
\par\end{centering}
\end{figure}

For illustration, Figure \ref{fig:Binomial Vs Uniform Distribution}
depicts the SA node degree distribution, obtained from a Monte Carlo
simulation of a shared channel, with $N_{s}=8640$ slots, adopting
random access, and uniform access, for various number of users. Figures
{[}a{]} through {[}d{]}, show the node degree distribution of $N_{u}=\left\{ 2,\:4,\:8,\:16\right\} $
users, respectively, with $n_{n_{u}}=4320$ symbols each. As shown
in the figures, for all the different number of users, the random
access schemes always show a binomial behavior in their simulated
degree distribution, that coincides perfectly with the derived analytical
distribution in (\ref{eq: Binomial distribution}). The plots visibly
show the non-zero probability of idle slots, whereas their exact values
are listed in the last row of Table \ref{tab:Capacity-Bounds-comparison},
for all the simulated users, as well as $N_{u}=\left\{ 10,\:12\right\} $.
On the other hand, as shown in the figures, for all the simulated
users, the degree distribution of uniform access is always constant.
It can be also noted that regardless of the number of users, for the
uniform access scheme, the probability of any channel slot being idle
is zero.

In the previous section, the spectral efficiency of random access,
was derived. Equation (\ref{eq: capacity_RA}) is solved numerically,
for the minimum $\frac{E_{b}}{N_{0}}$, required to achieve a spectral
efficiency, $\eta$. The minimum SNR per bit, $\left(\frac{E_{b}}{N_{0}}\right)_{RA}$,
of the random access scheme, is listed in the third row of Table \ref{tab:Capacity-Bounds-comparison},
for the same parameters as in Figure \ref{fig:Binomial Vs Uniform Distribution},
$N_{u}=\left\{ 2,\:4,\:8,\:10,\:12,\:16\right\} $ users, and rate-$\frac{1}{2}$
LDPC code. The previous section also derives the attainable spectral
efficiency of uniform access, such that (\ref{eq:capacity uniform})
gives a closed-form solution, that can be solved mathematically. The
minimum SNR per bit, $\left(\frac{E_{b}}{N_{0}}\right)_{UA}$, of
the uniform access scheme is calculated for the various number of
users, and listed in the fourth row in Table \ref{tab:Capacity-Bounds-comparison}.
Moreover, the fifth and sixth rows, respectively show the Sphere-Packing
Bound (SPB) \cite{shannon1967lower}, and the power efficiency gap
between the SPB and the uniform access. Furthermore, the seventh row
in the table, lists the gap in the power efficiency between the random
access and the uniform access schemes. As shown in the table, this
power efficiency gap goes hand in hand with the probability of idle
slots. 

\begin{table}[t]
\centering{}\caption{Capacity Bounds comparison of Random Access, Uniform Access, and Sphere-Packing
Bounds, for different spectral efficiencies, that correspond to different
number of users, with $n_{n_{u}}=4320$ symbols, and code rate, $r_{n_{u}}=\nicefrac{1}{2}$,
each, on a channel with $N_{s}=8640$ slots\label{tab:Capacity-Bounds-comparison}}
\begin{tabular}{|>{\centering}p{1.2cm}|>{\raggedright}m{0.5cm}|>{\centering}m{0.4cm}|>{\centering}m{0.4cm}|>{\centering}m{0.75cm}|>{\centering}m{0.75cm}|>{\centering}m{0.75cm}|>{\centering}m{0.75cm}|}
\hline 
$N_{u}$ & Single User & 2 & 4 & 8 & 10 & 12 & 16\tabularnewline
\hline 
\hline 
$\eta$ {[}b/s/Hz{]} & 1 & 1 & 2 & 4 & 5 & 6 & 8\tabularnewline
\hline 
$\left(\frac{E_{b}}{N_{0}}\right)_{RA}$ {[}dB{]} & 0 & 0.72 & 2.31 & 6.06 & 8.17 & 10.42 & 15.21\tabularnewline
\hline 
$\left(\frac{E_{b}}{N_{0}}\right)_{UA}$ {[}dB{]} & 0 & 0 & 1.76 & 5.74 & 7.92 & 10.21 & 15.03\tabularnewline
\hline 
$\left(\frac{E_{b}}{N_{0}}\right)_{SPB}$ {[}dB{]} & 0.69 & 0.69 & 2.49 & 6.59 & 8.84 & 11.20 & 16.17\tabularnewline
\hline 
$\Delta_{SPB-UA}$ {[}dB{]} & 0.69 & 0.69 & 0.73 & 0.85 & 0.92 & 0.99 & 1.13\tabularnewline
\hline 
$\Delta_{RA-UA}$ {[}dB{]} & 0 & 0.72 & 0.55 & 0.32 & 0.25 & 0.20 & 0.17\tabularnewline
\hline 
RA Prob. of empty slots & 0 & 0.25 & 0.06 & 3.9e-3 & 9.7e-4 & 2.4e-4 & 1.5e-5\tabularnewline
\hline 
\end{tabular}
\end{table}

In light of the above discussion, although random access schemes present
themselves as excellent candidates to meet the low-latency requirements,
the capacity bounds clearly indicate that uniform access schemes are
more power efficient. Consequently, this paper proposes a novel uniform
access scheme, dubbed Collision Diversity SCRAM (CoD-SCRAM), to leverage
the gap in capacity. Meanwhile, in order to maintain the low-latency,
the proposed scheme ensures that the coordination is kept minimal.

\section{LDPC: A Blast from the Past}

This section constitutes a crucial step in the proposal of the Collision-Diversity
SCRAM scheme. It highlights the analogy between the capacity-approaching
classical LDPC codes, and the proposed SCRAM algorithm. In essence,
it sheds light on the design aspects of powerful LDPC codes, and maps
and them to the SCRAM algorithm.

Inspired by BP, the essence of the SCRAM algorithm lies in the flow
of soft information on a joint three-layer Tanner graph, that comprises
variable nodes, and both SA and LDPC check nodes. Because the graphical
structure of the classical LDPC two-layer Tanner graph has a profound
impact on the performance of the LDPC code, the graphical features
of the three-layer Tanner graph of SCRAM are also provisioned to impact
its decoding performance drastically. These features are mainly portrayed
in cycles and trapping sets.

\subsection{Impact of Cycles}

An $(n,k)$ LDPC code is represented by a Tanner graph that comprises
$n$ variable nodes, and $m\ge n-k$ check nodes \cite{johnson2006introducing}.
The performance of the iterative BP algorithm is optimal provided
that the Tanner graph does not include cycles \cite{karimi2012message}.
The presence of short cycles in the Tanner graph hinders the performance
of BP and leads to error propagation \cite{xiao2019reed}.

On an LDPC Tanner graph, a cycle is defined as a path that starts
and terminates at the same node, and goes through a sequence of distinct
nodes via their connected edges \cite{karimi2012message}. A cycle
of length $L$ is referred to as an $L$- cycle. The girth of a Tanner
graph is denoted by $g$, and is defined as the length of the shortest
cycle within the graph \cite{li2015improved}. A cycle profile is
a quantitative analysis of the number of short cycles within a Tanner
graph, whose length is less than double the girth \cite{karimi2012message}.

Concerning SCRAM, two sub-definitions of cycles of the three-layer
Tanner graph are declared; a local cycle and a global cycle. A local
cycle on the SCRAM Tanner graph denotes a cycle that initiates and
terminates at a given variable node, that belongs to a certain user,
by means of traversing a path between variable nodes and LDPC check
nodes that represent the LDPC code of the denoted user. A global cycle
on the other hand, is defined as a cycle that initiates and terminates
at a variable node, that belongs to a certain user, by means of traversing
a path between variable nodes and LDPC check nodes that belong to
different users, via trespassing their commonly connected SA check
nodes.

The cycle-counting algorithm proposed in \cite{li2015improved} yields
the cycle profile of classical LDPC codes. In order to obtain the
cycle profile of SCRAM, two possible approaches could be followed.
The first approach involves the mapping of the SCRAM three-layer Tanner
graph to a joint SCRAM parity check matrix, which is then fed to the
cycle-counting algorithm in \cite{li2015improved}, yielding the cycle
profile of SCRAM. Alternatively, a thorough analysis of the the SCRAM
three-layer Tanner graph lead to the proposal of a cycle-counting
algorithm that is specifically tailored for SCRAM. This algorithm
focuses on the global cycles of length eight as it was proven by the
analysis to constitute a lower bound on the girth of SCRAM.

\subsection{Impact of Trapping Sets}

Another important aspect that hinders the performance of LDPC is the
Trapping Sets. The concept of trapping sets emerged from the stopping
sets that deteriorate the performance of LDPC codes on the Binary
Erasure Channel (BEC) \cite{hashemi2015characterization}. In a nutshell,
a stopping set refers to the set of variable nodes, in the Tanner
graph of an LDPC code, that are still erased (possess erased bits),
after the iterative decoding. 

Similar to the stopping sets, the trapping sets, on the Binary Symmetric
Channel (BSC), or the AWGN channel, denote the set of variable nodes
that are connected to unsatisfied check nodes, after exhausting the
predefined iterations of the iterative decoder \cite{karimi2012efficient}.
In a similar fashion to BEC, these are the sets that trap the LLRs
within the induced subgraph, preventing the decoder from convergence.
While the cycles contribute in bouncing the belief (or LLR) of one
variable node back to itself, violating the concept of extrinsic information,
the presence of trapping sets enhances the weight of the invalid bounced-back
beliefs. In other words, the variable nodes become trapped in their
own misleading beliefs, causing the iterative decoding to fail, and
the error rate curves to saturate. 

Although the mapping between the trapping sets on the two-layer Tanner
graph of LDPC, and the three-layer Tanner graph of SCRAM, is not straightforward,
the analogy still holds. In simple terms, similar to the discussion
on local and global cycles, a subgraph on the SCRAM Tanner graph,
that involves variable nodes, SA check nodes, and LDPC check nodes,
which cause the beliefs to be trapped, could be regarded as a global
SCRAM trapping set. 

\subsection{Problem Statement, Objective, and Design Guidelines}

To sum up, in a similar analogy to BP decoding of classical LDPC codes
on a two-layer Tanner graph, the SCRAM mechanism jointly resolves
collisions and decodes the LDPC codewords on a hybrid three-layer
Tanner graph. The results in \cite{nafie2018scram} depict the drastic
outperformance of SCRAM over the sequential state-of-the-art decoders
presented in the literature. A glimpse on the theoretical capacity
bounds, highlights the superiority of uniform access over the random
access schemes, due to the zero probability of idle slots attributed
to the uniform access. Consequently, attempting to replace the random
access with uniform access within the SCRAM mechanism is provisioned
to further improve its performance.

A first intuitive idea is to adopt a uniform Sequential SCRAM mechanism.
The key concept is to distribute the modulated symbols of the $N_{u}$
users, evenly among the $N_{s}$ slots, such that for $n_{s}=1,\ldots,N_{s}$,
every SA check node, $s_{n_{s}}$, possesses the same degree, $d_{s_{n_{s}}}$.
The simplicity of the sequential access scheme lies in the systematic
mapping of the users\textquoteright{} symbols, sequentially to the
channel slots. In other words, The first user should be assigned the
first $n_{n_{u}}$ slots, to sequentially transmit its $n_{n_{u}}$
symbols. The second user should then take the next $n_{n_{u}}$ slots.
This process is repeated till all the $N_{s}$ slots are exhausted.
In a similar fashion multiple rounds should be made to accommodate
the remaining users. The process is terminated when all the users
are accommodated.

A global 8\textminus cycle on the SCRAM three-layer Tanner graph comprises
three main constituent components; two connected variable nodes that
belong to one user, two connected variable nodes that belong to another
user, and two shared SA check nodes. A careful inspection of the sequential
uniform access scheme reveals that this access scheme constitutes
a rich environment for this typical cycle pattern. 

In the sequential uniform access, two or more users fully overlap
in a sequential fashion. This means that every variable node within
one user set, overlaps with the variable node (or nodes) of the same
respective index within the other user (or users) set. As a result,
if two variable nodes within one user set are connected by one or
more LDPC check nodes, it is guaranteed that their corresponding variable
nodes within the other user (or users) set are also connected. Since
the sequential channel access ensures that these connected variable
nodes on the different users\textquoteright{} sides share the same
SA check nodes, the typical $8-$cycle pattern is guaranteed. In conclusion,
in order to reduce the number of global $8-$cycles, the typical cycle
pattern has to be disrupted.

In progression, a modified version of the sequential uniform access,
dubbed interleaved uniform access is proposed. The key idea is to
maintain the uniform degree distribution of the channel slots, while
significantly reducing the number of global $8-$cycles. The essence
of the proposed access scheme lies in the adoption of random interleavers,
that break the typical cycle structure of global $8-$cycles. This
means that while the users are still assigned sequential channel slots,
prior to transmission, each user passes its modulated symbols to a
random interleaver. The interleaved symbols are then transmitted sequentially
over the channel slots. Thus, although similar to the sequential access,
two or more users fully overlap over one of the channel subgraphs,
the order of their symbols is different. As a result, it is no longer
guaranteed that two connected variable nodes on one user set, would
collide on the SA slots, with two connected variable nodes on the
other user (or users) set.

A thorough investigation of the sequential uniform access and the
interleaved uniform access schemes, reveals that both schemes posses
common structure of the so-called global trapping sets. In both schemes,
each group of users fully overlap on a subset of the channel slots,
yielding the overall Tanner graph of the SCRAM system, as multiple
disjoint subgraphs. Each of these disjoint subgraphs constitutes a
global SCRAM trapping set, which fully traps the beliefs within the
subgraph, resembling the definition of conventional trapping sets
on LDPC Tanner graphs. Even in the case of the interleaved uniform
access, where each user interleaves their symbols before transmission,
each group of users is assigned the same set of channel slots as in
the case of the sequential access scheme. This means that, each group
of users fully overlap on one subgraph, only in a different order
of their transmitted symbols.

In the next section, a novel channel access scheme, dubbed Collision
Diversity SCRAM (CoD SCRAM), that tackles the problem of global trapping
sets within the three-layer Tanner graph, is proposed. The proposed
scheme combines the advantages of the previously proposed schemes,
while optimizing the global trapping set structure. This means that
the proposed CoD SCRAM scheme maintains the uniform degree distribution
of the channel slots, just like the sequential and the interleaved
access schemes. Moreover, as will be illustrated in the subsequent
sections, the proposed scheme possesses less global $8-$cycles than
the interleaved access scheme, which is provisioned to have far less
ones than the sequential access scheme. Furthermore, the proposed
scheme tackles the problem of the trapping disjoint subgraphs. 

The essence of the proposed CoD SCRAM scheme lies in broadening the
spectrum of the joint SCRAM decoder, allowing it to gather reliable
beliefs from all nodes within the three-layer Tanner graph. Unlike
the sequential and the interleaved access schemes, each user within
the proposed access scheme is encouraged to transmit their symbols
on multiple subgraphs. A user with poor conditions, would have their
symbols spanning the multiple subgraph, wrecking the chance of a complete
set of users being trapped. In other words, the proposed scheme leverages
diversity by means of allowing (for the same channel slot degree)
the maximum number of users to collide.

\section{CoD-SCRAM: Uniform Interleaved SCRAM with Collision Diversity}

In compliance with the conventional SCRAM scheme, the proposed Collision
Diversity system accommodates $N_{u}$ users, with $n_{n_{u}}$ symbols
each. The shared channel comprises $N_{s}$ slots, with $n_{s}$ denoting
the slot index. Moreover, it is assumed that the number of channel
slots, $N_{s}$, is an integer multiple, $N_{subgraphs}$, of the
number of symbols per user, $n_{n_{u}}$. This means that the shared
access channel can be regarded as $N_{subgraphs}$ subchannels, with
$n_{n_{u}}$ slots each. Unlike the sequential and the interleaved
uniform access schemes, where two or more users fully collide on each
subgraph, rendering the three-layer Tanner graph as $N_{subgraphs}$
disjoint subgraphs, the Collision Diversity scheme provides a full-span
access across all the $N_{subgraphs}$.

Each user, $U_{n_{u}}$, hops over the subgraphs in a cyclic fashion
at a fixed step size, in order to evenly distribute its $n_{n_{u}}$
symbols among the $N_{subgraphs}$ subgraphs. Unlike the interleaved
uniform access, where the data symbols of each user are being interleaved
prior to transmission, in the CoD SCRAM scheme, it is the indices
of the channel slots that are actually interleaved. Each set of $n_{n_{u}}$
channel slots within one subgraph is interleaved separately. This
condition ensures that when the users hop over the subgraphs with
the interleaved slots, they still spread their symbols equally across
the different subgraphs. Moreover, assuming that the number of users,
$N_{u}$, is an integer multiple of the number of subgraphs, $N_{subgraphs}$,
each group of $N_{subgraphs}$ users experience the same interleaved
version of the channel slots. The interleaving process changes from
one set of $N_{subgraphs}$ users to another.

In order to minimize coordination, the interleaving of the channel
slots within each subgraph, can be fulfilled by means of a Linear
Feedback Shift Register (LFSR) \cite{wang1988linear}, with a seed
that is a function of both the user set index, and the subgraph index.
The user set index is denoted by $n_{set}$, and refers to the index
of the set of $N_{subgraphs}$ users that experience the same interleaved
version of the channel slots. The total number of user sets, $N_{sets}$,
is thus given by the ratio between the total number of users, $N_{u}$,
and the number of subgraphs, $N_{subgraphs}$. As a result, the user
set index, $n_{set}$, goes from $n_{set}=1$ to $n_{set}=N_{sets}=N_{u}N_{subgraphs}$.
The subgraph index, on the other hand, is denoted by $n_{sub}$, such
that $n_{sub}=1,\ldots,N_{subgraphs}$.

For $n_{u}=1,\ldots,N_{u}$, each user, $U_{n_{u}}$, first calculates
the index of the user set, $n_{set}$, that it belongs to. Because
each user set comprises $N_{subgraphs}$ users, the user set index
is given by $n_{set}=\lceil\frac{n_{u}}{N_{subgraphs}}\rceil$. After
that, each user, $U_{n_{u}}$, activates $N_{subgraphs}$ different
LFSRs, each seeded by the user's user set index, $n_{set}$, and the
subgraph index, $n_{sub}$, of the corresponding subgraph, such that
$n_{sub}=1,\ldots,N_{subgraphs}$. For $n_{set}=1,\ldots,N_{sets}$,
and $n_{sub}=1,\ldots,N_{subraphs}$, let $\mathbf{\Pi}^{\left(n_{set},\:n_{sub}\right)}=\left[\pi_{1}^{\left(n_{set},\:n_{sub}\right)},\ldots,\pi_{n_{n_{u}}}^{\left(n_{set},\:n_{sub}\right)}\right]$,
be the vector of $n_{n_{u}}$ interleaved indices of subgraph, $n_{sub}$,
and users that belong to user set, $n_{set}$. 

For convenience, it is assumed that for $n_{sub}=1,\ldots,N_{subgraphs}$,
the respective interleaver vectors, $\mathbf{\Pi}^{\left(n_{set},\:n_{sub}\right)}$,
correspond to the output of the LFSRs, added to a shift value, so
that the interleaved indices could directly point to the slot indices,
not within the subgraph, but within the total set of $N_{s}$ channel
slots. Thus, while all the LFSRs of the $N_{subgraphs}$ interleavers
would output randomized indices between $1$ and $n_{n_{u}}$, for
$n_{sub}=1,\ldots,N_{subgraphs}$, the indices within $\mathbf{\Pi}^{\left(n_{set},\:n_{sub}\right)}$
include a shift by $\left(n_{sub}-1\right)\:n_{n_{u}}$. 

After interleaving the slots, each of the $N_{u}$ users has access
to $N_{subgraphs}$ interleaved vectors, of the $N_{subgraphs}$ subblocks
of the channel slots, with $n_{n_{u}}$ interleaved slots each. The
channel access takes place concurrently for all the users, such that
each user hops over the subgraphs in a cyclic fashion, and picks one
slot at a time at a fixed step size. Within every user set, each user
starts at a different subgraph, picks up a specific slot within the
interleaved version of the subgraph, and hops to the next subgraph.
When the user exhausts all the subgraphs, it hops back to the first
subgraph, and continues with the pickup process. 

Within each user set, the first user, $U_{n_{u}}$, such that $n_{u}=\left(n_{set}-1\right)\:N_{subgrahps}+1$,
starts at the first interleaved subgraph, $\mathbf{\Pi}^{\left(n_{set},\:1\right)}$,
and picks up the first interleaved slot, $\pi_{1}^{\left(n_{set},\:1\right)}$,
to transmit its first symbol. After that, it hops to the second interleaved
subgraph, $\mathbf{\Pi}^{\left(n_{set},\:2\right)}$, and transmits
its second modulated symbol on the second interleaved slot, $\pi_{2}^{\left(n_{set},\:2\right)}$.
The user continues hopping till it exhausts all the subgraphs. At
this point, the user hops back to the first interleaved subgraph,
$\mathbf{\Pi}^{\left(n_{set},\:1\right)}$, and picks the interleaved
slot, $\pi_{N_{subgraphs}+1}^{\left(n_{set},\:1\right)}$, at index
$N_{subgraphs}+1$, to transmit its $N_{subgraphs}^{\textrm{th}}$
modulated symbol. The process continues till the user transmits all
its $n_{n_{u}}$ modulated symbols.

The second user within the same set, $U_{n_{u}}$, such that $n_{u}=\left(n_{set}-1\right)\:N_{subgrahps}+2$,
repeats the same process, except that it starts at the second interleaved
subgraph, $\mathbf{\Pi}^{\left(n_{set},\:2\right)}$, and picks up
the first interleaved slot, $\pi_{1}^{\left(n_{set},\:2\right)}$,
to transmit its first modulated symbol. Here again, when the user
exhausts all the subgraphs, it hops back to the first interleaved
subgraphs, and continues till it transmits all its $n_{n_{u}}$ modulated
symbols. 

The same process is repeated in parallel, for all the different $N_{sets}$
sets of users, and all the $N_{subgraphs}$ users per user set. Each
user set experiences a different version of the $N_{subgraphs}$ interleaved
vectors, based on its user set index, $n_{set}$, while each user
starts at a different interleaved subgraph, $\mathbf{\Pi}^{\left(n_{set},\:n_{sub}\right)}$,
based on its user index, $n_{u}$. In general terms, for $n_{u}=1,\ldots,N_{u}$,
the starting interleaved subgraph of user $U_{n_{u}}$, is $\mathbf{\Pi}^{\left(n_{set},\:n_{sub}\right)}$,
such that $n_{set}=\lceil\frac{n_{u}}{N_{subgraphs}}\rceil$, and
$n_{sub}=\mod\left(n_{u}-1,N_{subgraphs}\right)+1$. For $i=1,\ldots,n_{n_{u}}$,
user $U_{n_{u}}$ sends its $i^{\textrm{th}}$ modulated symbol, on
the $i^{\textrm{th}}$ interleaved index, within the interleaved vectors
of its current user set, starting at its calculated $n_{sub}$, and
going with incrementing $i$, in increments of one (with mod operation
to go back to first subgraph when all subgraphs are exhausted). Consequently,
if $n_{s}^{\left(n_{u},\:i\right)}$ denotes the index of the channel
slot, that user $U_{n_{u}}$, sends its $i^{\textrm{th}}$ modulated
symbols on, for $n_{u}=1,\ldots,N_{u}$, and $i=1,\ldots,n_{n_{u}}$,
$n_{s}^{\left(n_{u},\:i\right)}$ is given by

\begin{equation}
n_{s}^{\left(n_{u},\:i\right)}=\pi_{i}^{\left(n_{set},\:\mod\left(n_{u}+i-2,\:N_{subgraphs}\right)+1\right)},\label{eq:CoD-SA slot index}
\end{equation}
where the R.H.S of (\ref{eq:CoD-SA slot index}) corresponds to the
$i^{\textrm{th}}$ interleaved index, within the interleaved vector,
$\mathbf{\Pi}^{\left(n_{set},\:n_{sub}\right)}$, such that $n_{sub}=\mod\left(n_{u}+i-2,N_{subgraphs}\right)+1$,
and $n_{set}=\lceil\frac{n_{u}}{N_{subgraphs}}\rceil$. 

The corresponding CoD three-layer Tanner graph can be generated by
connecting SA check node, $s_{n_{s}^{\left(n_{u},\:i\right)}}$, where
$n_{s}^{\left(n_{u},\:i\right)}$ is calculated as per (\ref{eq:CoD-SA slot index}),
to variable node, $v_{n_{v}^{\left(n_{u},\:i\right)}}$, such that
for $n_{u}=1,\ldots,N_{u}$, and $i=1,\ldots,n_{n_{u}}$, $n_{v}^{\left(n_{u},\:i\right)}$
is given by 

\begin{equation}
n_{v}^{\left(n_{u},\:i\right)}=i+\left(n_{u}-1\right)n_{n_{u}}.\label{eq:CoD- VN slot index}
\end{equation}

\begin{figure}[tb]
\begin{centering}
\includegraphics[width=1\columnwidth]{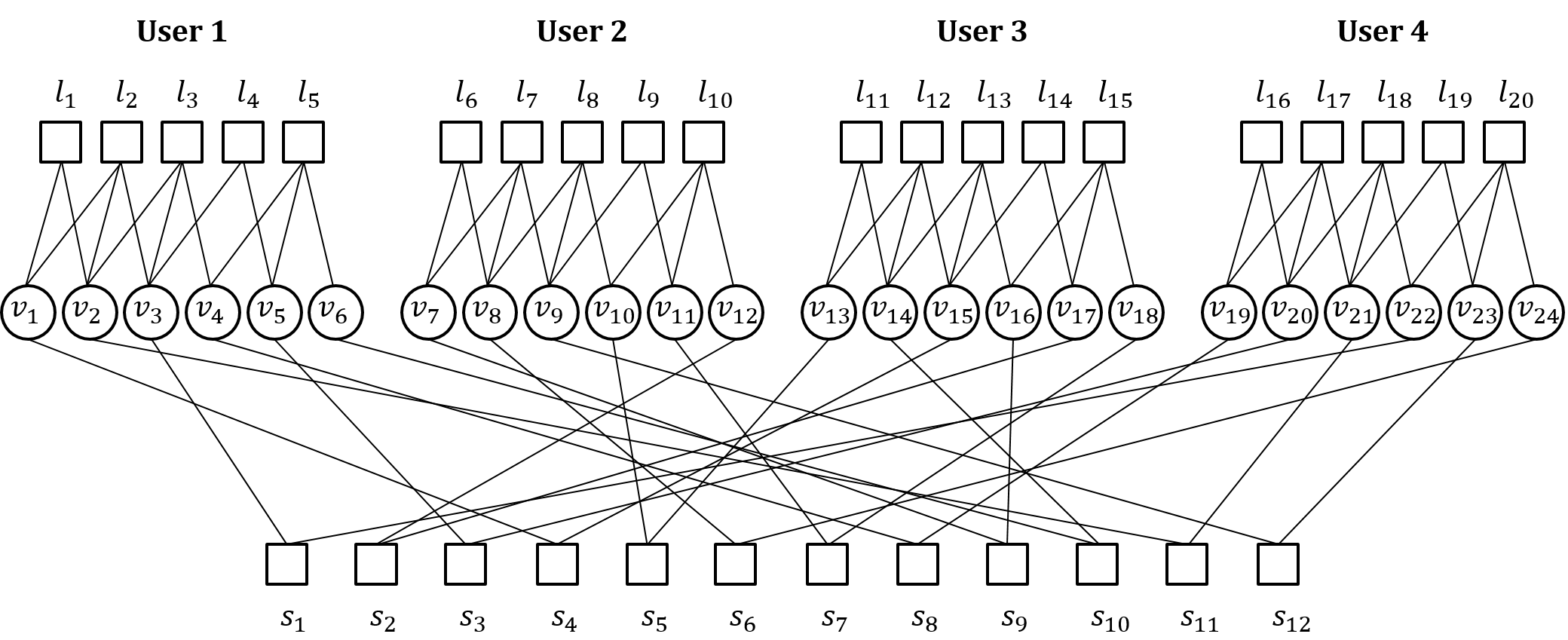}\caption{Interleaved Uniform Access with Collision Diversity example, of a
SCRAM system with $N_{u}=4$ users, $n_{n_{u}}=6$ modulated symbols
per user, and $N_{s}=12$ SA slots \label{fig:CoD Example}}
\par\end{centering}
\end{figure}

Consider the example in Figure \ref{fig:CoD Example}, of a CoD SCRAM
system with $N_{u}=4$ users, and $n_{n_{u}}=6$ symbols per user.
The shared channel comprises $N_{s}=12$ slots, that could be regarded
as $N_{subgraphs}=2$ subgraphs, with $n_{n_{u}}=6$ slots each. The
total number of user sets is given by $N_{sets}=\frac{N_{u}}{N_{subgraphs}}=2$,
with $N_{subgraphs}=2$ users per user set. The user set index, $n_{set}$,
is calculated as $n_{set}=\lceil\frac{n_{u}}{N_{subgraphs}}\rceil$,
such that users $U_{1}$ and $U_{2}$ belong to $n_{set}=1$, while
users $U_{3}$ and $U_{4}$ belong to $n_{set}=2$. Consequently,
users $U_{1}$ and $U_{2}$ generate $N_{subgraphs}=2$ interleaved
vectors, $\mathbf{\Pi}^{\left(1,\:1\right)}$ and $\mathbf{\Pi}^{\left(1,\:2\right)}$,
seeded by a function of $n_{set}=1$ and $n_{sub}=1$, and $2$, respectively.
Meanwhile, users $U_{3}$ and $U_{4}$ generate $N_{subgraphs}=2$
interleaved vectors, $\mathbf{\Pi}^{\left(2,\:1\right)}$ and $\mathbf{\Pi}^{\left(2,\:2\right)}$,
seeded by a function of $n_{set}=2$ and $n_{sub}=1$, and $2$, respectively. 

For the two interleaved instances of the second subgraph, i.e., $\mathbf{\Pi}^{\left(1,\:2\right)}$
and $\mathbf{\Pi}^{\left(2,\:2\right)}$, the output of the LFSRs
is shifted by $\left(n_{sub}-1\right)n_{n_{u}}$, such that $n_{sub}=2$.
This means that for both $n_{set}=1$ and $n_{set}=2$, the output
of the LFSR at $n_{sub}=2$, is shifted by six to point directly to
the corresponding index within the $N_{s}=12$ channel slots.

Let the two interleaved vectors at $n_{set}=1$, be $\mathbf{\Pi}^{\left(1,\:1\right)}=\left[4,\;6,\;1,\;5,\;3,\;2\right]$
and $\mathbf{\Pi}^{\left(1,\:2\right)}=\left[9,\;11,\;12,\;8,\;7,\;10\right]$,
which correspond to $n_{sub}=1$ and $n_{sub}=2$, respectively. Similarly,
for $n_{sub}=1$ and $n_{sub}=2$, let the two interleaved vectors
at $n_{set}=2$, be $\mathbf{\Pi}^{\left(2,\:1\right)}=\left[5,\;3,\;4,\;1,\;2,\;6\right]$
and $\mathbf{\Pi}^{\left(2,\:2\right)}=\left[8,\;10,\;11,\;9,\;12,\;7\right]$,
respectively. User $U_{1}$ belongs to user set $n_{set}=1$, and
starts at subgraph, $n_{sub}=\mod\left(n_{u}-1,N_{subgraphs}\right)+1=1$,
by picking up the first interleaved slot, $\pi_{1}^{\left(1,\:1\right)}=4$,
from $\mathbf{\Pi}^{\left(1,\:1\right)}$, to transmit its first modulated
symbol. After that, $U_{1}$ picks up the second interleaved slot,
$\pi_{2}^{\left(1,\:2\right)}=11$, from $\mathbf{\Pi}^{\left(1,\:2\right)}$,
to transmit its second modulated symbol. Now that all the subgraphs
are exhausted, $U_{1}$ goes back to the first subgraph and picks
up the third interleaved slot, $\pi_{3}^{\left(1,\:1\right)}=1$,
from $\mathbf{\Pi}^{\left(1,\:1\right)}$, to transmit its third modulated
symbol. $U_{1}$ keeps on hopping over the subgraphs till it transmits
its six modulated symbols. 

Similarly, user $U_{2}$ also belongs to user set $n_{set}=1$, but
starts at subgraph, $n_{sub}=2$, by picking up the first interleaved
slot, $\pi_{1}^{\left(1,\:2\right)}=9$, from $\mathbf{\Pi}^{\left(1,\:2\right)}$,
to transmit its first modulated symbol. User $U_{2}$ already exhausts
the subgraphs after the first pickup. As a result, for its next hop,
$U_{2}$ goes back to the first subgraph and picks up the second interleaved
slot, $\pi_{2}^{\left(1,\:1\right)}=6$, from $\mathbf{\Pi}^{\left(1,\:1\right)}$,
to transmit its second modulated symbol. $U_{2}$ then continues by
picking up the third interleaved slot, $\pi_{3}^{\left(1,\:2\right)}=12$,
from $\mathbf{\Pi}^{\left(1,\:2\right)}$, to transmit its third modulated
symbol, and so on. In a similar fashion, users $U_{3}$ and $U_{4}$
start at $n_{sub}=1$ and $n_{sub}=2$, respectively, at the interleaved
vectors, $\mathbf{\Pi}^{\left(2,\:1\right)}$ and $\mathbf{\Pi}^{\left(2,\:2\right)}$,
that correspond to $n_{set}=2$. 

Alternatively, (\ref{eq:CoD-SA slot index}) could directly compute
the slot index of the $i^{\textrm{th}}$ modulated symbol of user
$U_{n_{u}}$, for $n_{u}=1,\ldots,N_{u}$, and $i=1,\ldots,n_{n_{u}}$.
For example, the third modulated symbol of user $U_{1}$, corresponds
to $n_{u}=1$ and $i=3$. According to (\ref{eq:CoD-SA slot index}),
the corresponding slot index, $n_{s}^{\left(1,\:3\right)}$, is given
by $\pi_{3}^{\left(n_{set},\:1\right)}$, such that $n_{set}=1$ is
the user set index of $U_{1}$. Thus, the slot index at which $U_{1}$
sends its third modulated symbol is $\pi_{3}^{\left(1,\:1\right)}=1$,
which corresponds to the third interleaved index, on the first subgraph,
of the first interleaving set, $\mathbf{\Pi}^{\left(1,\:1\right)}$.
According to (\ref{eq:CoD- VN slot index}), the corresponding variable
node index, $n_{v}^{\left(1,\:3\right)}$, of the third modulated
symbol of $U_{1}$, is given by $n_{v}^{\left(1,\:3\right)}=3$. Therefore,
on the three-layer Tanner graph, variable node $v_{3}$, is connected
to SA check node $s_{1}$.

\begin{figure}[tbh]
\begin{centering}
\includegraphics[width=1\columnwidth]{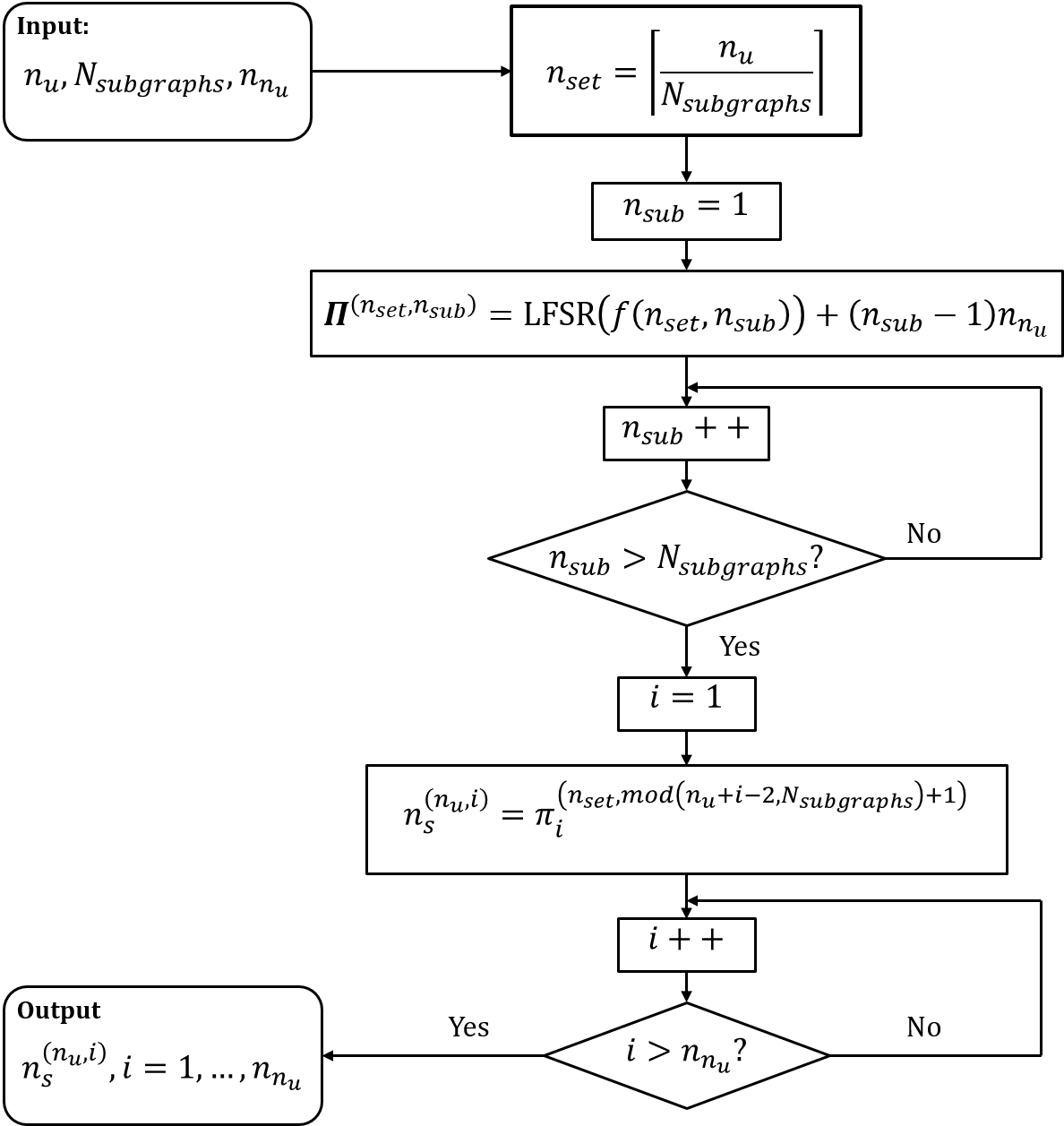}\caption{Flowchart of Interleaved Uniform Access SCRAM with Collision Diversity,
on each user terminal, $U_{n_{u}}$, in the uplink scenario\label{fig:CoD Flowchart}}
\par\end{centering}
\end{figure}

Figure \ref{fig:CoD Flowchart} shows the flowchart of the proposed
CoD SCRAM, at an arbitrary user terminal, $U_{n_{u}}$. Based on the
assigned user ID, $n_{u}$, the user calculates the index of the user
set, $n_{set}$, to which it belongs. The user then loops over the
indices, $n_{sub}$, of the $N_{subgraphs}$ blocks of channel slots.
For $n_{sub}=1,\ldots,N_{subgraphs}$, the user generates $N_{subgraphs}$
interleaved vectors, $\mathbf{\Pi}^{\left(n_{set},\:n_{sub}\right)}$,
generated by LFSRs, that are seeded by a function of both $n_{set}$
and $n_{sub}$, and shifted to match the index of the slots within
the $N_{s}$ slots. After that, the user loops over its $n_{n_{u}}$
symbols, and for every symbol index $i$, such that $i=1,\ldots,n_{n_{u}}$
, $U_{n_{u}}$, calculates the corresponding slot index, $n_{s}^{\left(n_{u},\:i\right)}$.
The algorithm halts when the corresponding slots of all the $n_{n_{u}}$
symbols are calculated.

\section{Cycle Profile of CoD-SCRAM}

In this section, the cycle profile of the proposed CoD SCRAM, is to
be analyzed. The analysis is carried out in two parallel directions.
First, the three-layer Tanner graph of the collision diversity SCRAM
scheme is mapped to a two-dimensional hybrid matrix, which is then
fed to the cycle counting algorithm in \cite{li2015improved}, yielding
the cycle-profile of the proposed scheme. Meanwhile, the global $8-$cycles
counter algorithm, quantifies the number of added global $8-$cycles,
that result from the adoption of the collision diversity scheme.

In order to have a full picture, the analysis of the SCRAM systems
with sequential, interleaved, and random access schemes is included
for convenience. The SCRAM system accommodates $N_{u}=4$ user, with
$n_{n_{u}}=4320$ symbols each. The adopted LDPC encoders are assumed
to be identical for all the four users, and correspond to the $(4320,2160)$
code, taken from the DVB-NGH standard \cite{gomez2014dvb}. The shared
access channel comprises $N_{s}=8640$ slots, which can be regarded
as $N_{subgraphs}=2$ subgraphs, with $n_{n_{u}}=4320$ slots each.

\begin{table}[b]
\caption{Cycle Profile of SCRAM, with $N_{u}=4$ users, adopting the $(4320,2160)$
DVB-NGH LDPC code, with Interleaved Uniform Access with Collision
Diversity, on a channel with $N_{s}=8640$ slots\label{tab:Cycle Profile}}

\centering{}%
\begin{tabular}{|>{\centering}p{1.5cm}|>{\centering}p{0.7cm}|>{\centering}p{0.8cm}|>{\centering}p{0.9cm}|>{\centering}p{0.7cm}|>{\centering}m{0.8cm}|}
\hline 
\multirow{2}{1.5cm}{} & \multicolumn{2}{>{\centering}p{1.5cm}|}{Cycle Profile (LDPC)} & Global $8-$Cycles & \multicolumn{2}{>{\centering}p{1.5cm}|}{Cycle Profile (SCRAM)}\tabularnewline
\cline{2-6} \cline{3-6} \cline{4-6} \cline{5-6} \cline{6-6} 
 & $C_{6}$ & $C_{8}$ &  & \multirow{1}{0.7cm}{$C_{6}$} & $C_{8}$\tabularnewline
\hline 
DVB NGH LDPC & 31200 & 1558340 &  &  & \tabularnewline
\hline 
Collision Diversity &  &  & 482 & 124800 & 6233842\tabularnewline
\hline 
Sequential  &  &  & 138224 & 124800 & 6371584\tabularnewline
\hline 
Interleaved  &  &  & 1002 & 124800 & 6234362\tabularnewline
\hline 
Random Access &  &  & 725 & 124800 & 6234085\tabularnewline
\hline 
\end{tabular}
\end{table}

The cycle profile of the SCRAM systems with the different access methods,
along with the number of global $8-$cycles is depicted in Table \ref{tab:Cycle Profile}.
The first row shows the cycle profile of the $(4320,2160)$ DVB-NGH
\cite{gomez2014dvb} LDPC code. The second row shows the cycle profile
of the proposed CoD SCRAM, along with its global $8-$cycle count.
The results of the sequential, interleaved, and random access schemes,
are depicted in the third, fourth, and fifth rows, respectively.

For all the access schemes, the number of $6-$cycles from the cycle
profile is four times that of pure LDPC. This proves that all the
$6-$cycles in the three-layer Tanner graph of the respective SCRAM
system are local LDPC cycles, and that the schemes do not add global
cycles of length six. Concerning the $8-$cycles, for all the access
methods, the total number of $8-$cycles from the cycle profile corresponds
to four times that of LDPC, in addition to the number of global $8-$cycles,
obtained from the global $8-$cycles counter. As a result, the validity
of the algorithmic global $8-$cycles counter is verified. 

As shown in the table, the CoD scheme possesses the least number of
global $8-$cycles, in comparison to the other access schemes. The
small number of global $8-$cycles, along with its ability to unlock
the trapping sets, are provisioned to let the CoD scheme outperform
the other tackled access schemes. 

\section{Results}

This section is dedicated to evaluating the PER performance of the
proposed CoD SCRAM scheme, in comparison to the sequential, interleaved,
and random access SCRAM schemes. For all the access schemes, the system
incorporates $N_{u}=4$ users. Each user adopts a $(4320,2160)$,
rate $\nicefrac{1}{2}$, LDPC code, taken from the DVB-NGH standard
\cite{gomez2014dvb}. The $n_{n_{u}}=4320$ encoded bits of each user
are first BPSK modulated, and then equalized by the estimated magnitude
of the Rayleigh fading coefficients, assuming perfect CSI at the transmitter.
The shared access channel comprises $N_{s}=8640$ frequency subcarriers.
The superimposed received symbols are perturbed by complex AWGN of
zero mean, and variance $N_{0}$. At the receiver side, the joint
SCRAM decoder runs 50 decoding iterations on the three-layer Tanner
graph.

\begin{figure}[b]
\begin{centering}
\includegraphics[width=0.95\columnwidth]{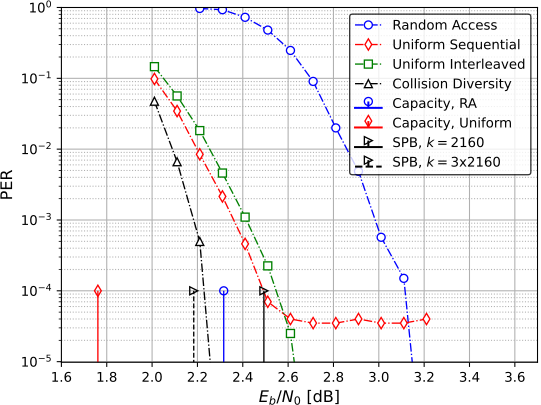}\caption{PER comparison of Interleaved Uniform Access with Collision Diversity,
Interleaved Uniform Access, Sequential Uniform Access, and Random
Access vs. SPB, of SCRAM with $N_{u}=4$ Users, $N_{s}=8640$ slots,
and $(4320,2160)$ NGH LDPC, over Rayleigh fading and pre-equalization
at the transmitter side, simulated for a sample size of $10^{5}$
packets per user\label{fig:PER-CoD}}
\par\end{centering}
\end{figure}

Figure \ref{fig:PER-CoD} shows the PER performance, of the proposed
CoD SCRAM scheme, simulated vs. $\frac{E_{b}}{N_{0}}$. The figure
also depicts the results of the sequential, interleaved, and the random
access SCRAM schemes, along with the capacity bounds of both uniform
and random access. As provisioned, the proposed collision diversity
scheme drastically outperforms the sequential, interleaved, and the
random access schemes. As depicted in the figure, at the PER of $10^{-3}$
and $10^{-4}$, the CoD scheme outperforms the interleaved uniform
access by approximately $0.23\:\textrm{dB}$ and $0.32\:\textrm{dB}$,
respectively. The collision diversity scheme is also $0.096\:\textrm{dB}$
better than the capacity bounds of random access. 

In comparison to the capacity of uniform access, the CoD scheme is
yet $0.46\:\textrm{dB}$ short of the ultimate capacity bound. The
reason behind this capacity gap, is the relatively short information
block length $(2160)$, of the adopted LDPC code. This in fact leads
to the inclusion of the Sphere-Packing Bound (SPB) \cite{shannon1967lower,dolinar1998code},
which is a more realistic bound, that takes the information block
length into consideration. The figure also shows the SPB calculated
for $k=2160$, and $k=3\times2160$, plotted in a solid and a dashed
line, respectively. As shown in the figure, the proposed CoD SCRAM
is superior to the SPB calculated at $k=2160$ by approximately $0.25\:\textrm{dB}$.
The noted outperformance of the proposed CoD SCRAM can be justified
by a virtual enhancement in the effective block length.

More specifically, a CoD SCRAM system with $N_{u}$ users over $N_{s}=N_{subgraphs}\times n$
channel slots, ensures that each user collides with $(N_{subgraphs}-1)(N_{u}N_{subgraphs})$
users. Consequently, by adopting the law of large numbers \cite{yao2015law},
the effective information block length is enhanced by a factor of
approximately $((N_{subgraphs}-1)(N_{u}N_{subgraphs})+1)$. For the
simulations presented herein, with $N_{u}=4$ users, $N_{s}=2n$,
the number of subgraphs is given by $N_{subgraphs}=2$. Consequently,
the effective block length is given by $3k$. For this reason, Figure
\ref{fig:PER-CoD} shows the SPB calculated at block length of $3k$.
As shown in the figure, the performance of the CoD SCRAM is only approximately
$0.056\:\textrm{dB}$ away from that SPB.

In order to highlight the essence of the proposed CoD SCRAM scheme,
the simulations are extended to higher number of users, at the same
channel load. In this simulation apparatus, the objective is to assess
the impact of the proposed CoD scheme on the effective information
block length. As a result, for the adopted simulations, a target channel
load, $D$, is set and kept fixed. Meanwhile, the simulations are
carried out for various number of users, $N_{u}$. In order to keep
the channel load fixed, going for higher number of users corresponds
to increasing the number of channel slots, $N_{s}$, in order to meet
the target channel load.

\begin{figure}[tbh]
\begin{centering}
\includegraphics[width=0.95\columnwidth]{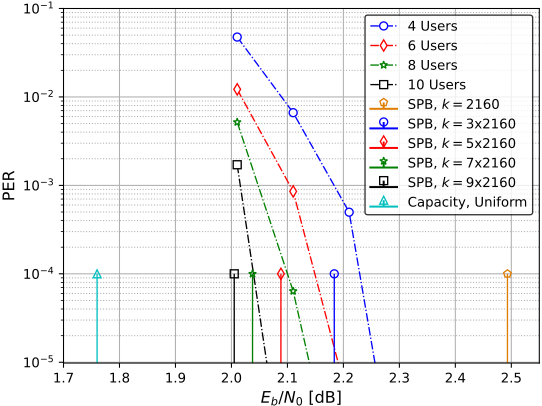}\caption{PER comparison of Interleaved Uniform Access with Collision Diversity,
for $N_{u}=\left\{ 4,6,8,10\right\} $ users, channel load of 1, and
$(4320,2160)$ NGH LDPC, over Rayleigh fading and pre-equalization
at the transmitter side, simulated for a sample size of $10^{5}$
packets per user, plotted against SPB at different information block
lengths\label{fig:PER-CoD-D_1}}
\par\end{centering}
\end{figure}

Figure \ref{fig:PER-CoD-D_1} depicts the PER performance of the proposed
CoD SCRAM system at a target channel load of $D=1$. The system is
simulated for $N_{u}=\left\{ 4,6,8,10\right\} $ users. Thus, the
number of adopted channel slots is given by $N_{s}=\left\{ 2n,3n,4n,5n\right\} $,
respectively, where $n$ is the length of the LDPC codeword at the
output of the LDPC encoder. This means that for the CoD scheme, the
number of subgraphs is given by $N_{subgraphs}=\left\{ 2,3,4,5\right\} $,
respectively. Similar to the previous simulation, the users adopt
the rate $\nicefrac{1}{2}$, $(4320,2160)$ LDPC code from the DVB
NGH standard \cite{gomez2014dvb}.

Figure \ref{fig:PER-CoD-D_1} also depicts the theoretical capacity
bounds. The leftmost bound corresponds to the ultimate capacity bound,
while the rightmost bound represents the SPB calculated at information
block length $k=2160$. Meanwhile, the SPB at the enhanced length
of $3k$, $5k$, $7k$, and $9k$, that corresponds to $N_{u}=4$,
$6$, $8$, and $10$ users, respectively, is calculated and plotted
for convenience. The figure sheds the light on numerous observations.
First, although all the plotted PER curves correspond to a fixed channel
load, $D$, going for higher number of users leads to further improvement
in the PER performance. For example, at the PER of $10^{-3}$, the
performance of the proposed CoD SCRAM system with six, eight, and
ten users, is superior to that of the four users by approximately
$0.08\:\textrm{dB}$, $0.14\:\textrm{dB}$, and $0.17\:\textrm{dB}$,
respectively. In comparison to the ultimate capacity bound, the figure
shows that at the PER of $10^{-4}$, the gap between the performance
of the four, six, eight, and ten users, to the ultimate capacity bound,
reduces to approximately $0.46\:\textrm{dB}$, $0.38\:\textrm{dB}$,
$0.34\:\textrm{dB}$, and $0.27\:\textrm{dB}$, respectively.

In comparison to the SPB, the figure shows that all the plotted curves
are superior to the SPB, calculated at $k=2160$. As shown in the
figure, at the PER of $10^{-4}$, the CoD SCRAM system with $N_{u}=4$,
$6$, $8$, and $10$ users outperforms the SPB at $k=2160$ by approximately,
$0.27\:\textrm{dB}$, $0.35\:\textrm{dB}$, $0.39\:\textrm{dB}$,
and $0.46\:\textrm{dB}$ respectively. As a result, not only does
the proposed CoD improve the effective information block length, but
also accommodating higher number of users leads to a better utilization
of the granted inherent coordination that reflects on the respective
PER performance.

Finally, in order to assess the quantification of the effective information
block length enhancement, for the CoD systems with $N_{u}=4$, $6$,
$8$, and $10$ users, and $N_{subgraphs}=2$, $3$, $4$, and $5$,
respectively, the SPB is calculated at a block length of $3k$, $5k$,
$7k$, and $9k$, respectively, where $k=2160$ is the information
block length of the underlying LDPC code. By comparing the PER curves
to the respective bounds, it can be shown that the gap is subtle.
For instance, at the PER of $10^{-4}$, the gap between the PER curves
of the proposed CoD system with $N_{u}=4$, $6$, $8$, and $10$
users, and the SPB bounds at a block length of $3k$, $5k$, $7k$,
and $9k$, is approximately $0.04\:\textrm{dB}$, $0.06\:\textrm{dB}$,
$0.07\:\textrm{dB}$, and $0.025\:\textrm{dB}$, respectively.

\section{Conclusions}

In this paper, a novel Non-Orthogonal Multiple Access (NOMA)-based
approach dubbed Collision Diversity (CoD) SCRAM is proposed. The proposed
CoD scheme is regarded as an enhancement of the conventional Slotted
Coded Random Access Multiplexing (SCRAM) mechanism. The essence of
SCRAM lies in its ability to jointly resolve collisions and decodes
the Low Density Parity Check (LDPC) codewords, in a similar analogy
to Belief Propagation (BP) on a joint three-layer Tanner graph. The
propsed CoD SCRAM is pillared on three main terminologies. First,
it adopts an information-theoretic approach in order to maximize the
attainable Spectral Efficiency. Secondly, it utilizes the analogy
between the two-layer Tanner graph of classical LDPC codes, and the
three-layer Tanner graph of SCRAM, in order to adopt the well-established
design tools of powerful LDPC codes to enhance the graphical structure
of the SCRAM three-layer Tanner graph. Finally, the proposed CoD SCRAM
leverages the collisions among the transmitted packets in order to
enhance the respective information block length. 

\appendices{}

\bibliographystyle{IEEEtran}
\bibliography{literature}

\end{document}